\documentstyle[tighten,preprint,aps,prd]{revtex}
\begin{document}
\draft

\hyphenation{
mani-fold
mani-folds
}


\def\Bbb{\bf}

\def\BbbR{{\Bbb R}}
\def\BbbZ{{\Bbb Z}}

\def\sutwo{{\hbox{${\rm SU}(2)$}}}
\def\sothree{{\hbox{${\rm SO}(3)$}}}

\def\undertn{{\rlap{\lower1.8ex\hbox{$\tilde{\phantom{n}}$}}n}}


\preprint{\vbox{\baselineskip=12pt
\rightline{WISC-MILW-94-TH-20}
\rightline{gr-qc/9408033}}}
\title{Chern-Simons functional and the no-boundary proposal in \\
Bianchi~IX quantum cosmology}
\author{Jorma Louko\cite{jorma}}
\address{
Department of Physics,
University of
Wisconsin--Milwaukee,
\\
P.O.\ Box 413,
Milwaukee, Wisconsin 53201, USA}
\date{Revised version, October 1994}
\maketitle
\begin{abstract}%
The Chern-Simons functional $S_{\rm CS}$ is an exact solution to the
Ashtekar-Hamilton-Jacobi equation of general relativity with a nonzero
cosmological constant. In this paper we consider $S_{\rm CS}$ in Bianchi type
IX cosmology with $S^3$ spatial surfaces. We show that among the classical
solutions generated by~$S_{\rm CS}$, there is a two-parameter family of
Euclidean spacetimes that have a regular NUT-type closing.
When two of the three scale factors are equal, these spacetimes reduce to a
one-parameter family within the Euclidean Taub-NUT-de~Sitter metrics. For a
nonzero cosmological constant, $\exp(iS_{\rm CS})$ therefore provides a
semiclassical estimate to the Bianchi~IX no-boundary wave function in
Ashtekar's variables.
\end{abstract}

\narrowtext

\section{Introduction}

Kodama\cite{kodama} has pointed out that the Chern-Simons functional $S_{\rm
CS}$ provides an exact solution to the Ashtekar-Hamilton-Jacobi equation of
general relativity with a nonzero cosmological constant. One therefore
anticipates that when the theory is canonically quantized in the connection
representation, the quantum constraint equations should possess a solution of
the form $\exp(iS_{\rm CS})$, either exactly or in some approximate
semiclassical sense\cite{kodama,bgp,changsoo}.

Investigating this anticipation within full general relativity faces the
difficult issues of regularization\cite{bgp,changsoo}, and even in
minisuperspace models there emerge issues of factor ordering\cite{kodama,mena}.
Further, even if a state of the form $\exp(iS_{\rm CS})$ is shown to satisfy
the quantum constraints in the sense of (functional) differential equations,
one may ask whether this state is normalizable with respect to any inner
product that is proposed to define the physical Hilbert space of the
theory\cite{mena}. Nevertheless, leaving for now such questions aside, one is
led to identify a state of the form $\exp(iS_{\rm CS})$ as corresponding, in a
semiclassical sense, to the family of classical solutions to the Einstein
equations that are generated by regarding $S_{\rm CS}$ as a particular solution
to the Ashtekar-Hamilton-Jacobi equation. It is therefore of interest to
understand what the properties of this family of classical spacetimes are. For
earlier work relating to $S_{\rm CS}$, see
Refs.\cite{kodama,bgp,changsoo,mena,samuel,AAbook}.
For a supersymmetric generalization, see Ref.\cite{shiraishi}.

The purpose of the present paper is to investigate $S_{\rm CS}$ in the Bianchi
type IX spatially homogeneous cosmological model\cite{ryan,jantzen12} with
$S^3$ spatial surfaces. We shall show that among the classical solutions
generated by~$S_{\rm CS}$, there is a two-parameter family of Euclidean
spacetimes that have a regular closing of the NUT-type\cite{nuts1,nuts2}. This
implies that, in this model, a wave function of the semiclassical form
$\exp(iS_{\rm CS})$ can be regarded as compatible with the no-boundary proposal
of Hartle and Hawking\cite{vatican,HH,hawNB}. We shall also note in passing
that when two of the three scale factors are equal, the Euclidean spacetimes
corresponding to $S_{\rm CS}$ reduce to a one-parameter family within the
Taub-NUT-de~Sitter metrics\cite{carter,exact,jlr} and can be given in closed
form.

We now proceed to prove these assertions. Except when otherwise stated, the
notation follows that of Ref.\cite{kodama}.

\section{The model}

The (Lorentzian) spacetime metric is
\begin{equation}
ds^2 = {1\over 32\pi^2}
\left[
- \sigma_1 \sigma_2 \sigma_3 \,
{\undertn}^2 dt^2
+ \sigma_1^{-1} \sigma_2 \sigma_3 {(\chi^1)}^2
+ \sigma_2^{-1} \sigma_3 \sigma_1 {(\chi^2)}^2
+ \sigma_3^{-1} \sigma_1 \sigma_2 {(\chi^3)}^2
\right]
\ \ ,
\label{metric}
\end{equation}
where $\chi^I$ are the left invariant one-forms on $\sutwo\simeq S^3$
satisfying $d\chi^I = \case{1}{2} \epsilon_{IJK} \chi^J \wedge \chi^K$. The
rescaled lapse $\undertn$ and the components $\sigma_I$ of the inverse
densitized triad are functions of $t$ only. The Ashtekar action\cite{AAbook}
takes the form\cite{kodama}
\begin{equation}
S = \int dt \left( - \sigma_I {\dot A}_I - \undertn h \right)
\ \ ,
\label{action}
\end{equation}
where a sum over the repeated index is understood. The Hamiltonian constraint
$h(\sigma_I,A_I)$ is given by
\begin{equation}
h =
- \sigma_1 \sigma_2 \left( A_1 A_2 \mp i A_3 \right)
- \sigma_2 \sigma_3 \left( A_2 A_3 \mp i A_1 \right)
- \sigma_3 \sigma_1 \left( A_3 A_1 \mp i A_2 \right)
+ 3\lambda \sigma_1 \sigma_2 \sigma_3
\ \ .
\label{hamiltonian}
\end{equation}
Here $\lambda=\Lambda/(96\pi^2)$, where $\Lambda$ is the cosmological constant.
We shall assume throughout $\lambda\ne0$. The upper and lower signs correspond
to the two possible signs of $i$ in the definition of the Ashtekar connection;
we refer to Ref.\cite{kodama} for the details. Note that $A_I$ differs by an
overall factor of $i$ from the conventions adopted in Ref.\cite{AAbook}.

{}From Eq.~(\ref{action}), the fundamental Poisson brackets are $\left\{A_I,
\sigma_J \right\}= - \delta_{IJ}$. Thus, if $A_I$ is regarded as a coordinate,
its conjugate momentum is $-\sigma_I$. A na\"{\i}ve counting indicates that the
general solution to the equations of motion contains four constants of
integration. It appears unknown whether the global structure of the
(Lorentzian) solution space is consistent with the existence of such four
constants\cite{grubisic}; we shall, however, not attempt to address these
global issues at the level of the present paper.

Dirac quantization\cite{diracbook} with $A_I$ as the configuration variable
leads to the Wheeler-DeWitt~-type equation
\begin{equation}
{\hat h}\left( i(\partial/\partial A_I) , A_I \right) \psi = 0
\label{wdw}
\end{equation}
for the wave function $\psi(A_I)$. We shall not attempt to discuss the factor
ordering in this equation, nor the choice of an inner product\cite{AAbook}.

Given a solution to the Wheeler-DeWitt equation~(\ref{wdw}) with the
(approximate) form~$\exp(iS)$, a semiclassical expansion\cite{kiefer} shows
that $S$ (approximately) solves the Hamilton-Jacobi equation
\begin{equation}
h \left( -(\partial S/\partial A_I) , A_I \right) = 0
\ \ .
\label{hj}
\end{equation}
The wave function is therefore, through semiclassical correspondence,
associated with the spacetimes obtained by solving the equations
\begin{mathletters}
\label{hjeom}
\begin{eqnarray}
\sigma_I &=& - {\partial S \over \partial A_I}
\ \ ,
\label{hjeom1}
\\
\noalign{\smallskip}
{\dot A}_I &=& {\undertn} \left\{ A_I, h \right\}
= - {\undertn} {\partial h \over \partial \sigma_I}
\ \ .
\label{hjeom2}
\end{eqnarray}
\end{mathletters}%
By standard Hamilton-Jacobi theory\cite{ll,goldstein,halljeru}, the solutions
to (\ref{hjeom}) for the given $S$ are a two-parameter family of solutions to
the classical equations of motion.

The Chern-Simons functional $S_{\rm CS}$ takes the
form\cite{kodama}\footnote{We have chosen the overall sign in Eq.~(\ref{cs}) to
differ from that adopted in Ref.\cite{kodama}. This reflects our using wave
functions of the form $\exp(iS)$ rather than~$\exp(-iS)$.}
\begin{equation}
S_{\rm CS} = - {1\over\lambda} \left[ A_1 A_2 A_3
\mp {i\over2}\left( A_1^2 + A_2^2 + A_3^2 \right)
\right]
\ \ .
\label{cs}
\end{equation}
It is readily seen that $S_{\rm CS}$ solves the Hamilton-Jacobi
equation~(\ref{hj}). We are now interested in the corresponding classical
solutions.

Inserting $S_{\rm CS}$ (\ref{cs}) into~(\ref{hjeom}), one obtains after some
rearrangement the equations
\begin{mathletters}
\label{cseom}
\begin{equation}
\begin{array}{rcl}
\lambda \sigma_1 &=& A_2 A_3 \mp iA_1
\ \ ,
\\
\lambda \sigma_2 &=& A_3 A_1 \mp iA_2
\ \ ,
\\
\lambda \sigma_3 &=& A_1 A_2 \mp iA_3
\ \ ,
\end{array}
\label{cseom1}
\end{equation}
and
\begin{equation}
\begin{array}{rcl}
{\dot A}_1 &=& - \lambda {\undertn} \sigma_2 \sigma_3
\ \ ,
\\
{\dot A}_2 &=& - \lambda {\undertn} \sigma_3 \sigma_1
\ \ ,
\\
{\dot A}_3 &=& - \lambda {\undertn} \sigma_1 \sigma_2
\ \ .
\end{array}
\label{cseom2}
\end{equation}
\end{mathletters}%
The solutions to Eqs.~(\ref{cseom}) are (anti-)self-dual metrics, and can
therefore not generically be Lorentzian\cite{kodama,samuel}. We shall seek
Euclidean solutions.

We take $\sigma_I$ to be real. Requiring the spatial metric in
Eq.~(\ref{metric}) to be positive definite implies that either $\sigma_I$ are
all positive, or one of them is positive and two are negative. As
Eqs.~(\ref{cseom}) are invariant under a simultaneous sign change in any two of
the $\sigma_I$ and the corresponding two $A_I$, we can without loss of
generality take all $\sigma_I$ positive. We now set ${\undertn} = \pm i
{(\sigma_1 \sigma_2 \sigma_3)}^{-1/2}$, which makes $t$ proportional to the
Euclidean proper time. We also rewrite $\sigma_I$ in terms of positive-valued
scale factors $a$, $b$, and~$c$, defined by
\begin{equation}
\begin{array}{rcl}
\sigma_1^{-1} \sigma_2 \sigma_3 &=& \case{1}{4} a^2
\ \ ,
\\
\noalign{\smallskip}
\sigma_2^{-1} \sigma_3 \sigma_1 &=& \case{1}{4} b^2
\ \ ,
\\
\noalign{\smallskip}
\sigma_3^{-1} \sigma_1 \sigma_2 &=& \case{1}{4} c^2
\ \ .
\end{array}
\end{equation}
Note that at the isotropic limit $a=b=c$, the constant $t$ surfaces in the
metric (\ref{metric}) are round three-spheres with sectional curvature
$32\pi^2/(a^2)$. Writing finally $A_I=\pm i B_I$, equations (\ref{cseom}) take
the form
\begin{mathletters}
\label{cseeom}
\begin{equation}
\begin{array}{rcl}
\case{1}{4}\lambda bc &=& B_1 - B_2 B_3
\ \ ,
\\
\noalign{\smallskip}
\case{1}{4}\lambda ca &=& B_2 - B_3 B_1
\ \ ,
\\
\noalign{\smallskip}
\case{1}{4}\lambda ab &=& B_3 - B_1 B_2
\ \ ,
\end{array}
\label{cseeom1}
\end{equation}
and
\begin{equation}
\begin{array}{rcl}
{\dot B}_1 &=& - \case{1}{2} \lambda a
\ \ ,
\\
\noalign{\smallskip}
{\dot B}_2 &=& - \case{1}{2} \lambda b
\ \ ,
\\
\noalign{\smallskip}
{\dot B}_3 &=& - \case{1}{2} \lambda c
\ \ .
\end{array}
\label{cseeom2}
\end{equation}
\end{mathletters}%

At the isotropic limit, $a=b=c$, Eqs.~(\ref{cseeom}) imply $B_1=B_2=B_3$, and
the solution is easily found\cite{kodama}: the metric is just that of Euclidean
space of constant sectional curvature~$\Lambda/3$, which can be understood as a
Euclidean section of (anti-)de~Sitter space for $\Lambda>0$ ($\Lambda<0$). We
shall now consider solutions that are near the isotropic limit. For this
purpose we introduce the Misner-type parameters\cite{misner}
\begin{mathletters}
\begin{equation}
\begin{array}{rcl}
a &=& \exp\left(\alpha + \beta^+ + \sqrt{3} \beta^-\right)
\ \ ,
\\
\noalign{\smallskip}
b &=& \exp\left(\alpha + \beta^+ - \sqrt{3} \beta^-\right)
\ \ ,
\\
\noalign{\smallskip}
c &=& \exp\left(\alpha - 2\beta^+\right)
\ \ ,
\end{array}
\end{equation}
and
\begin{equation}
\begin{array}{rcl}
B_1 &=& \case{1}{2} F \exp\left(g^+ + \sqrt{3} g^-\right)
\ \ ,
\\
\noalign{\smallskip}
B_2 &=& \case{1}{2} F \exp\left(g^+ - \sqrt{3} g^-\right)
\ \ ,
\\
\noalign{\smallskip}
B_3 &=& \case{1}{2} F \exp\left(-2 g^+\right)
\ \ ,
\end{array}
\end{equation}
\end{mathletters}%
and expand Eqs.~(\ref{cseeom}) to linear order in the anisotropy variables
$\beta^\pm$ and~$g^\pm$.
One obtains
\begin{mathletters}
\label{isoeom}
\begin{eqnarray}
{\dot F} &=& -\lambda e^\alpha
\ \ ,
\\
\lambda e^{2\alpha} &=& F(2-F)
\ \ ,
\\
{d\over dt} \left(F g^\pm\right) &=& -\lambda e^\alpha \beta^\pm
\ \ ,
\\
\lambda e^{2\alpha}\beta^\pm &=& -F(2+F)g^\pm
\ \ .
\end{eqnarray}
\end{mathletters}%

For $\lambda>0$, the solution to Eqs.~(\ref{isoeom}) is
\begin{mathletters}
\label{linsol}
\begin{eqnarray}
e^\alpha &=& {1\over \sqrt{\lambda}} \sin(\sqrt{\lambda} t)
\ \ ,
\\
F &=& 1 + \cos(\sqrt{\lambda} t)
\ \ ,
\\
\beta^\pm &=& \beta_0^\pm
\left( 2 + \tan^2( \case{1}{2} \sqrt{\lambda} t ) \right)
\tan^2( \case{1}{2} \sqrt{\lambda} t )
\ \ ,
\label{linsol3}
\\
g^\pm &=& - \beta_0^\pm \tan^4( \case{1}{2} \sqrt{\lambda} t )
\ \ ,
\label{linsol4}
\end{eqnarray}
\end{mathletters}%
where $\beta_0^\pm$ are constants and the range of $t$ is
$0<t<\pi/\sqrt\lambda$.
For $\lambda<0$, there are two solutions. One is given by the formulas
(\ref{linsol}), understood in the sense of analytic continuation in~$\lambda$;
the range of $t$ is $0<t<\infty$. The other solution is
\begin{mathletters}
\label{2linsol}
\begin{eqnarray}
e^\alpha &=& {-1\over \sqrt{-\lambda}} \sin(\sqrt{-\lambda} t)
\ \ ,
\\
F &=& 1 - \cosh(\sqrt{-\lambda} t)
\ \ ,
\\
\beta^\pm &=& \gamma_0^\pm
\left(\coth^2( \case{1}{2} \sqrt{-\lambda} t ) -2 \right)
\coth^2( \case{1}{2} \sqrt{-\lambda} t )
\ \ ,
\label{2linsol3}
\\
g^\pm &=& - \gamma_0^\pm \coth^4( \case{1}{2} \sqrt{-\lambda} t )
\ \ ,
\label{2linsol4}
\end{eqnarray}
\end{mathletters}%
where $\gamma_0^\pm$ are constants and the range of $t$ is $-\infty<t<0$.

When $\beta_0^+=\beta_0^-=0$, the linearized solution (\ref{linsol}) reduces
for both signs of $\lambda$ to the exact isotropic solution mentioned above.
When $\beta_0^\pm$ are not both equal to zero, the linearized solution
(\ref{linsol}) for $\lambda>0$ grows out of the domain of validity of the
linearized equations at large values of~$\sqrt{\lambda}t$, and the same is true
for $\lambda<0$ when $|\beta_0^\pm|$ are not much smaller than unity. However,
for any $\lambda\ne0$ and any values of~$\beta_0^\pm$, the linearized solution
(\ref{linsol}) is in the domain of validity of the linearized equations for
sufficiently small~$t$, and further it becomes asymptotically accurate as
$t\to0$. We infer that there exists a two-parameter family of solutions to the
exact equations (\ref{cseeom}), such that these exact solutions are well
approximated by (\ref{linsol}) at the limit $t\to0$.

When $\gamma_0^+=\gamma_0^-=0$, the second linearized solution (\ref{2linsol})
reduces to the exact isotropic solution mentioned above, with the coordinate
time $t$ now running in the opposite direction compared with the
$\beta_0^\pm=0$ limit of~(\ref{linsol}). When $\gamma_0^\pm$ are not both equal
to zero, the linearized solution (\ref{2linsol}) is within the domain of
validity of the linearized equations only when $|\gamma_0^\pm|\ll1$
and~$|\sqrt{-\lambda}t|\gg1$. This linearized solution will not be important
for our conclusions.

Recall now that we began by defining the metrics, both Lorentzian and
Euclidean, in terms of the 3+1 split expression~(\ref{metric}). It can now be
verified that inserting the linearized solution (\ref{linsol}) into
(\ref{metric}) gives a Euclidean metric that can be regularly extended to $t=0$
by adding just one point to the manifold: one can view the new point as the
coordinate singularity at the origin of a hyperspherical coordinate system in
which $t$ is the radial coordinate. The crucial fact that makes this extension
possible is the $O(t^2)$ suppression of the anisotropy (\ref{linsol3}) as
$t\to0$. Since $t\to0$ is the limit where the linearized solution is accurate,
we see that the corresponding exact metrics can be similarly extended to $t=0$.
In the terminology of Refs.\cite{nuts1,nuts2}, the closing of the geometry at
$t=0$ is of the NUT-type.

This regular closing of the geometry is precisely the property characterizing
the classical solutions that are relevant for the no-boundary proposal of
Hartle and Hawking\cite{vatican,HH,hawNB}, in the sense that wave functions
satisfying the no-boundary proposal are expected to get their dominant
semiclassical contribution from one or more such regular classical
solutions\cite{halljeru,hallhartle,jjhjl3}. We therefore conclude that in our
model, a wave function of the semiclassical form $\exp(iS_{\rm CS})$ is
compatible with a semiclassical estimate to the no-boundary wave function.

\section{Discussion}

We have investigated the Chern-Simons functional $S_{\rm CS}$ as a particular
solution to the Ashtekar-Hamilton-Jabobi equation in the Bianchi type~IX
cosmological model with $S^3$ spatial surfaces and a nonzero cosmological
constant. We showed that among the classical solutions generated by~$S_{\rm
CS}$, there is a two-parameter family of Euclidean spacetimes that have a
regular NUT-type closing. Hence, in this model, a wave function of the
semiclassical form $\exp(iS_{\rm CS})$ in the connection representation of
Ashtekar's variables is compatible with a semiclassical estimate to the
no-boundary wave function of Hartle and Hawking. Several comments are now in
order.

(1). Recall that we chose the spatial surfaces to be $\sutwo\simeq S^3$.
Another possible choice compatible with the  Bianchi~IX homogeneity type is
$\sutwo/{\BbbZ_2}\simeq\sothree\simeq\BbbR P^3$. In this case the Euclidean
solutions corresponding to the $t\to0$ limit of (\ref{linsol}) are not
regularly extendible. However, they can be extended into spaces that have an
orbifold-type singularity\cite{orbifold}. It has been suggested that the
no-boundary proposal could be meaningfully broadened to include such
geometries\cite{conifold}.

(2). At the limit $\lambda\to0$, $S_{\rm CS}$ (\ref{cs}) diverges. This is in
agreement with the obstacles discussed in Ref.\cite{loukoholo} to obtaining a
semiclassical estimate to the no-boundary wave function in Ashtekar's variables
for a vanishing cosmological constant. The crux of the problem is that without
a cosmological constant, the Hamiltonian constraint~(\ref{hamiltonian}) reduces
to a pure kinetic term.

(3). At the limit $\lambda\to0$, the linearized solutions (\ref{linsol}) and
(\ref{2linsol}) yield small anisotropy approximations to self-dual vacuum
solutions, provided that before taking the limit the constants $\beta_0^\pm$
are made proportional to $\lambda^{-1}$ and the constants $\gamma_0^\pm$
proportional to~$\lambda^2$. In the terminology of Ref.\cite{nuts1},
(\ref{linsol}) yields an approximation to metrics with
$\lambda_1=\lambda_2=\lambda_3=1$, and (\ref{2linsol}) yields an approximation
to metrics with $\lambda_1=\lambda_2=\lambda_3=0$.

(4). In our minisuperspace model, the gauge freedom pertaining to the Gauss and
diffeomorphism constraints\cite{AAbook} was fixed to begin with, and one can
view the expression (\ref{cs}) in essence as the definition of the Chern-Simons
functional. In more general contexts one may however raise questions regarding
the behavior of the Chern-Simons functional under large gauge transformations.
For these issues, see Ref.\cite{AAbook}.

(5). In the special case where two of the three scale factors are equal, the
classical Euclidean solutions corresponding to $S_{\rm CS}$ can be found in
closed form. To see this, we return to Eqs.~(\ref{cseom}), taking again
$\sigma_I$ to be positive. We now set $\sigma_1=\sigma_2$ and ${\undertn}= \pm
i \sigma_1^{-2}$, and write again $A_I=\pm i B_I$. The equations then imply
$B_1=B_2$ and reduce to
\begin{mathletters}
\label{tneeom}
\begin{equation}
\begin{array}{rcl}
\lambda \sigma_1 &=& B_1(1 - B_3)
\ \ ,
\\
\lambda \sigma_3 &=& B_3 - B_1^2
\ \ ,
\end{array}
\label{tneeom1}
\end{equation}
and
\begin{equation}
\begin{array}{rcl}
{\dot B}_1 &=& - \lambda \sigma_3 \sigma_1^{-1}
\ \ ,
\\
\noalign{\smallskip}
{\dot B}_3 &=& - \lambda
\ \ .
\end{array}
\end{equation}
\end{mathletters}%
The solution is straightforward to find. The connection is given by
\begin{equation}
\begin{array}{rcl}
B_1^2 &=& 1- 2\lambda t + k t^2
\ \ ,
\\
\noalign{\smallskip}
B_3 &=& 1- \lambda t
\ \ ,
\end{array}
\end{equation}
where $k$ is an integration constant and the additive
constant in $t$ has been chosen in a convenient way.
$\sigma_I$ are then obtained from~(\ref{tneeom1}).
Writing finally $t=32\pi^2\rho$ and $k=-f\lambda/(32\pi^2)$,
the metric (\ref{metric}) takes the form
\begin{eqnarray}
ds^2 &=&
{(1+f\rho) d\rho^2
\over \rho \left[ 1 - (2\Lambda/3)\rho - (f\Lambda/3)\rho^2 \right] }
+
{ \rho \left[ 1 - (2\Lambda/3)\rho -
(f\Lambda/3)\rho^2 \right] \over (1+f\rho)
}
\, {(\chi^3)}^2
\nonumber
\\
\noalign{\vskip2\jot}
&&+
\rho (1+f\rho)
\left( {(\chi^1)}^2 + {(\chi^2)}^2 \right)
\ \ ,
\label{tnds}
\end{eqnarray}
which is recognized as a one-parameter family within the two-parameter set of
Euclidean Taub-NUT-de~Sitter metrics\cite{carter,exact,jlr}. The range of
$\rho$ in which (\ref{tnds}) gives a positive definite metric depends on
$\Lambda$ and~$f$ but includes always an open (possibly semi-infinite) interval
whose lower end is at $\rho=0$, where the geometry has a regular NUT-type
closing\cite{nuts1,nuts2,jlr}: this corresponds to the linearized
solution~(\ref{linsol}). A metric corresponding to the second linearized
solution (\ref{2linsol}) is recovered from (\ref{tnds}) with $\Lambda<0$ and
$f\approx-\Lambda/3$ at large negative values of~$\rho$. The isotropic solution
is obtained with $f=-\Lambda/3$, for either sign of $\Lambda$, and the
anisotropic solution found in Ref.\cite{kodama} is obtained with $f=
\Lambda/3$.

(6). In the Taub truncation of the Bianchi~IX model, in which two of the three
scale factors are set equal to begin with, the Hamilton-Jacobi function
generating the one-parameter family of solutions (\ref{tnds}) in the
conventional metric variables can be given in terms of elementary
functions\cite{jlr}. Its form (for a nonvanishing $\Lambda$) is considerably
more cumbersome than that of the corresponding Chern-Simons functional obtained
from~(\ref{cs}).

(7). Finally, one must ask whether the connection between the Chern-Simons
functional and the no-boundary proposal in Bianchi type~IX might reflect more
general phenomena. We leave this question a subject to future work.

\acknowledgments
I would like to thank Abhay Ashtekar, Akio Hosoya, and Hideo Kodama for
discussions, and Chris Isham and Gary Gibbons for their hospitality at the
London Mathematical Society meeting ``Quantum Concepts in Space and Time"
(Durham, UK, July 1994), where this work was initiated. This work was supported
in part by the NSF grant PHY91-05935.

\end{document}